# Electrical, elastic properties and defect structures of isotactic polypropylene composites doped with nanographite and graphene nanoparticles

L.V. Elnikova[1*], A.N. Ozerin[2], V.G. Shevchenko[2], A.T. Ponomarenko[2], P.M. Nedorezova[3], O.M. Palaznik[3], V.V. Skoi[4,5], A.I. Kuklin[4,5]

[1]NRC "Kurchatov Institute" – Alikhanov Institute for Theoretical and Experimental Physics, 117218, Moscow, Russia,
[2]Enikolopov Institute of Synthetic Polymeric Materials, RAS, 117393, Moscow, Russia,
[3]Semenov Federal Research Center for Chemical Physics, RAS, 119991, Moscow, Russia
[4]Joint Institute for Nuclear Research, 141980, Dubna, Russia
[5]Moscow Institute of Physics and Technology, 141701, Dolgoprudny, Russia

[*]Corresponding author: elnikova@itep.ru

**Abstract**
Conducting polymers have wide technological applications in sensors, actuators, electric and optical devices, solar cells *etc*. To improve their operational performance, mechanical, thermal, electrical and optical properties, such polymers are doped with carbon allotrope nanofillers. Functionality of the novel nanocomposite polymers may be stipulated by size characteristics of nanoparticles and the polymer, different physical effects like charge transfer in such objects *etc*. We characterize and analyze structure, elastic, electric properties and of novel polymer nanocomposites, isotactic polypropylene (iPP) with high crystallinity, doped with graphene nanoplates (GNP) and nanographite particles at different concentrations and sizes about 100 nm, basing on the results of dynamic mechanical analysis (DMA), dielectric spectroscopy, small-angle neutron scattering (SANS) and theoretical modeling. Carbon NPs aggregated in fractal objects in the bulk of iPP change its mechanical plastic, elastic and electric properties comparing with pristine polymer. We study modification of nanofiller morphology with the concept of Cosserat elasticity which involves description of the behavior of linear topological defects caused aggregation of nanographite and GNPs. We supply our experimental data with numerical simulations on the lattice in frames of the model of Cosserat elasticity to estimate some mechanical characteristics of the whole composite iPP.

**Key words**: iPP/GNP composites, iPP/nanographite composites, SANS, dielectric spectroscopy, Cosserat elasticity theory, disclinations

## 1. Introduction

Polymer nanocomposites filled with carbon allotropes have wide engineering applications in optoelectronics, photonics, sensing materials and actuators and other fields of human activity [1, 2]. The functional properties of polymer change when carbon allotropes are added [3,4,5], changes in morphology and size effects lead to changes in electrical and mechanical characteristics; doping of polymer with carbon nanofillers is accompanied by modification of surface, changes in polymer structure in the bulk and interactions between nanofillers, formation of defect structures there [6].

For isotactic polypropylene (iPP) doped with carbon allotrope nanofillers (nanographites, GNPs *etc*.) there are results on dynamic mechanical analysis (DMA) [7,8,9,10], dielectric spectroscopy, atomic force microscopy (AFM), differential scanning calorimetry (DSC), scanning electron microscopy (SEM), Raman scattering, small-angle X-ray scattering [3,4,5] and small-angle neutron scattering (SANS) [11].

Dynamic mechanical analysis (DMA) for the iPP-graphene systems [7] revealed variations of the storage ($E'$) and loss (tgδ) modulii with increasing of contents of graphene nanoparticles and temperature. Three relaxation processes (α, β, γ) at the $E'$ and tgδ curves are possible for such composites (β relaxation is associated with generalized motion in the amorphous regions during the glass transition, γ address to the mobility of crystallites).

From dielectric spectroscopy for the iPP-graphene systems combined with X-ray and DMA studies in the appropriate concentration range of NGP's fillers (by 4.5 vol.%) in iPP the authors found decreasing of dielectric permittivity and dielectric losses and the growth of crystallization point of the polymer with increase the filler content [8,9,10]. The elasticity modulus of the iPP composite increased by 25-35% when adding 1-1.5 vol.% of GNP's [9].

The SANS method provides nondestructive structure analysis in the length scale from 1 Å and explicitly characterizes the morphology of nanofillers in the bulk of the material [12]. The SANS data exhibited fractally modified geometry and carbon NP's sizes in the bulk of iPP [11].

Modification of geometry of the allotrope nanofillers in iPP is associated with appearance of defects, for instance, point defects, disclinations and dislocations.

Recently [13], for theoretical explanation of nanofillers (graphene and nanographite) aggregation, the Cosserat model of elasticity was grounded. The Cosserat concept [14,15] is a gauge theory, which differs from known elasticity theories by the expansion of the gauge field created by rotation, thereby creating additional degrees of freedom in the material. In context of similar modern problems, the Cosserat theory has been developed in [16,17,18,19,20,21,22] and references therein.

Using the fracton representation [13,18,19,20], the relations between defect structures, elastisity and electromagnetic properties were developed [22,23]; in particular, it is possible to describe percolations phenomena in polymer networks [21].

For the structurally related polymer systems [16], the isotropic elastic constants were defined in terms of the Cosserat theory comparing with experimental results.

The number of authors considers topological defects disclinations in appearing at formation of pentagons and heptagons instead of regular hexagon graphene lattice, point defects and grain boundaries [17] and references therein.

In our manuscript, we gather an experimental reports giving rise to development the theoretical approach to explanation of aggregation and self-organization of nanographite and graphene nanofillers in the polymer iPP matrix, and we provide modeling of the electrical, elastic and mechanical properties of these composite systems.

2. **Experimental**
2.1. **Sample preparation**

The samples of isotactic polypropylene (iPP) are filled by *n situ* polymerization with GNP at concentrations of 0.7 and 1.8 wt%, and nanographite at 1.5 and 3.6 wt%.

The chemical formula of iPP is $(C_3H_6)_n$, its density is 0.9–0.91 g/cm$^3$, and the degree of crystallinity is 60%. Nanographite particles are in the form of plates with diameter 112.7 nm and thickness of 47.3 nm.

Graphene nanoplates were produced by chemical or thermal reduction of graphite oxide (TRGO) [24]. Graphite oxide was produced using modified method of Hammers - oxidizing graphite by $KMnO_4$ in concentrated $H_2SO_4$.

After X-ray diffraction analysis [4] of GNP and TRGO powders performed using ADP-1 diffractometer, the authors [4] calculated the values of crystallite size 1.127 and 1.003 nm for GNP and TRGO, respectively. Synthesized few-layer particles contain 3-5 layers of graphene. The approximate dimensions of individual GNP particle are 100 nm × 100 nm × 1.127 nm.

Synthesis of nanocomposites was done in bulk propylene, as described in [4,5]. The metallocene catalyst was used in the synthesis of composites is highly active and isospecific in propylene polymerization, producing iPP of high molecular weight [25].

Concentration of filler in composites was varied by changing polymerization time. Final product was unloaded from the reactor, washed successively by a mixture of ethyl alcohol and HCl (10% solution), ethyl alcohol and then dried in vacuum at 60°C until constant weight.

Test specimen were cut from films 100–300 mm thick, pressure molded at 190°C and pressure 10 MPa at cooling rate 16 K·min$^{-1}$ [6, 7].

## 2.2. DMA, dielectric spectroscopy and mechanical properties

For the iPP-graphene systems [7] significant increase in Young's modulus is demonstrated, as GNP is incorporated in the iPP polymeric matrix, from 1280 ± 42 MPa in neat iPP to approximately 1920 ± 63 MPa in the iPP-graphene nanocomposite with the highest graphene content (17.4 wt.%), i.e. 50 % gain in rigidity.

The data on DMA and other structural methods [3,8,9,10] for different ordered carbon allotrope structures, including GNP, in the syndiotactic PP matrix in the temperature range from –60 to 160°C shown temperature decreasing of $E'$, temperature and concentration dependent peaks of mechanical and dielectric losses, as well as for graphene nanifillers (by 2.6wt%) in the iPP matrix.

In [4], the stress-strain properties of iPP/GNP composites with nanofiller content from 0.1 to 9.71 wt.% were studied upon quasistatic tension. Adding a small filler content (2-3 wt.%) in the polymer matrix leads to an increase matrix in the elasticity modulus upon extension by 25–35% compared to the initial polymer. After ultrasonic treatment of iPP/GNP samples, the relative elongation upon breakage and yield points vary differently depending on GNP's content.

Dielectric spectroscopy of iPP/GNP composites [4] confirmed a significantly increase of dielectric losses and the dielectric permeability of the composite in the range concentration of GNP's 0.1-9.71 wt.% at a frequency of 4.8 GHz. Adding of nanofillers GNP to iPP influences on the percolation threshold, it is reached at 2.1–4.2 wt.% of GNP nanofillers, which is lower than for graphite [4, 9]. Also iPP/GNP composites demonstrate high values of dielectric permeability ε' and dielectric losses ε", which grow with an increase of polymer filling.

## 2.3. Small-angle neutron scattering

For SANS measurements [11], we used the YuMO spectrometer at the IBR2 reactor in Dubna, Russian Federation [26], the neutron wavelength is $\lambda$= 0.7–6Å; the neutron flux on a sample is ~$10^7$ n/(s×cm$^2$) [27], diameter of neutron beam on the sample is 14 mm. The solid film-like specimens with different nanofillers (iPP/GNP, iPP/nanographite) were fixed in the holder and put into thermo box. The thickness values for the specimens were normalized to thickness of iPP film 368 μm.

In the SANS measurements with YuMO, we recorded counts versus time of flight from the 16 rings of two detectors. Recalculation and normalization count using the gauge standard of the known cross section *vs* time of flight to the differential scattering cross section $d\Sigma/d\Omega(Q)$ and normalization on the sample thickness was realized by program SAS [28].

To analyze the experimental small-angle scattering curves, we used a number of the following procedures of the ATSAS 2.4 software package [29].

Preliminary processing of the initial scattering curves and registration of scattering by the reference sample was performed using the PRIMUS procedure of ATSAS [29]. As the reference scattering, which was subtracted from the experimental curve of small-angle scattering of samples *I(Q)*, scattering from a sample of matrix polymer (iPP) was used. Thus, after taking into account reference scattering, the experimental small-angle scattering curves are characterized by scattering from only heterogeneous regions ("scattering particles") in the system having a scattering length different from the scattering length of the polymer matrix.

We used the GNOM procedure of ATSAS to calculate regularized scattering curves $I_{reg}(Q)$, optimized over the entire range of scattering angles, the particle distribution function, the integral values of the inertia radii of the particles of the scattering phase and the particle size distribution.

The experimental values of the intensity $I(Q)$ of SANS and the regularized SAXS (X-ray) curves $I_{reg}(Q)$ calculated under the GNOM procedure for samples nanographite 1.8, 3.6 wt% and GNP 0.7, 1.8 wt%, excluding scattering from the reference sample iPP, Fig. 1.

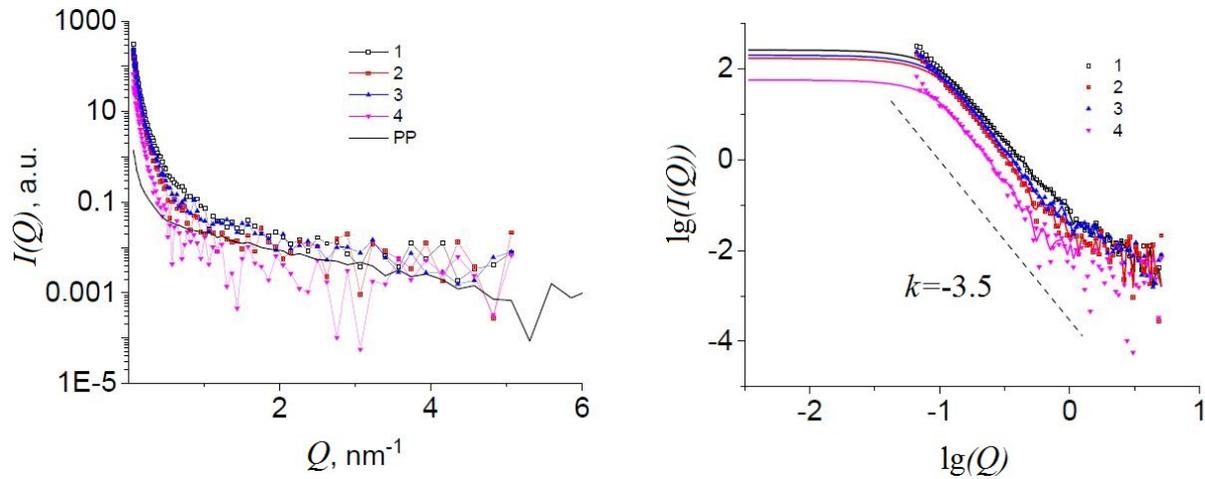

Fig. 1. The experimental SANS intensity $I(Q)$ of the samples 1 – nanographite 3.6 wt%; 2 – nanographite 1.5 wt%; 3 – GNP 1.8 wt%; 4 – GNP 0.7 wt%. The solid lines correspond to the regularized $I_{reg}(Q)$ curves. The value $k$ means the slope of the linear sections of the scattering curves and defines fractal dimensionality [11].

The scattering particles have an almost identical spatial structure and their scattering pattern corresponds to scattering by a physical fractal object with dimension $d_s = 6 - |k| = 2.5$, corresponding to a surface fractal. I.e. we get the dense compact aggregated particles with a rugged surface. The upper size range of these physical fractals exceeds the spatial resolution of the small angle neutron scattering method $L_{max} = 2\pi/Q_{min} = 94$ nm, implemented in the experiments of this work ($Q_{min} = 0.0665$ nm$^{-1}$).

Since the GNP and nanographite samples used in the work contain particles in the form of plates, the regularized scattering curves from these samples were primarily used to calculate the distribution function of particle thicknesses under the assumption of a polydisperse system of plate-shaped particles with thickness $T$ (the distance distribution function of thickness, assuming a polydisperse system of flat particles). The calculation results are shown in Fig. 2.

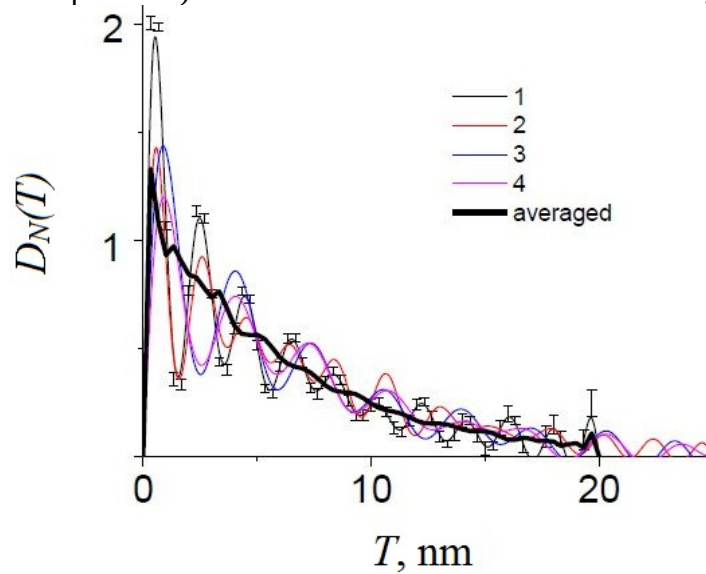

Fig. 2. Normalized distribution functions of particle thicknesses $D_N$ (thin lines) under the assumption of a polydisperse system of plate-shaped particles with a thickness *T* calculated from the scattering curves for GNP and nanographite particles. The thick line is the smoothed curve, with adjacent averaging, weighted average value [11].

Similar to the scattering curves, the distribution functions for the GNP and nanographite samples turned out to be identical to each other. The system of scattering particles in the GNP and nanographite samples is characterized by high polydispersity. The particle size distribution contains wafers with a thickness of 1 to 20 nm, whereas the thickness of the initial plates is 47.3 nm and 1 nm for nanographite and GNP samples, respectively.

The radius of gyration $R_t = T^2/12$ of the particle thickness determined from the slope of the linear part of the $\ln(Q^2 I(Q))$-$Q^2$ plot (the Guinier plot) in the reciprocal space and that calculated by the indirect transform method [12] applied to the whole experimental scattering curve while using the GNOM procedure, were close to each other and equal, on average, to 5.5 nm.

The particle size distribution contains wafers with a thickness of 1 to 20 nm, whereas the thickness of the initial plates is 47.3 nm and 1 nm for nanographite and GNP samples, respectively.

3. **Modeling**

Elasticity of a 2*D* graphene crystalline honeycomb lattice, in the commonly used membrane theory, is defined by topological defects dislocations with their translation Burgers vector $\vec{b} = n\vec{a}_1 + m\vec{a}_2$ (which is a topological invariant), where $\vec{a}_{12} = (3d_{cc}/2 \pm \sqrt{3}d_{cc}/2)$, $d_{cc}$ = 1.42 Å [17], the nearest-neighbor interatomic distance in graphene, $\vec{a}$ is the vector connecting with disclinations there, the pair of integers (*n, m*) is addressed to dislocations. Such defects form pentagons and heptagons in the graphene lattice. The second topological invariant is a microorientation angle θ. The core of the shortest Burgers vector dislocation (1,0) $|\vec{b}_{(1,0)}| = \sqrt{3}d_{cc}$ = 2.46 Å, contains an edge-sharing heptagon-pentagon pair, its Burgers vector is oriented along the zigzag direction. The type of dislocations is (1,1) dislocation, its Burgers vector equals = 4.23 Å and inserts a semi-infinite strip along the zigzag direction of graphene. The core of the dislocation with the same Burgers vector can be constructed from two $|\vec{b}_{(1,0)}|$ =2.46 Å dislocations: (1,0) and (0,1) [17]. Grain boundaries are considered as periodic arrays of dislocations.

Defects dislocations and disclinations demonstrate restrict mobility, dislocations obey glide constraint, they are moving along their Burgers vector, provided that total number of lattice sites is conserved, while disclinations cannot move without creating dislocations. Similar properties are observed in type-I gapless fracton phases.

The densities of crystalline defects vacansies, dislocations and disclinations are given by $\rho_{vac} = \partial_i u^i$, $\rho_{disc} = \varepsilon^{kl} \partial_k \partial_l u^i_{sing}$, $\rho_{discl} = \varepsilon^{ij} \partial_i \partial_j \phi$ [18].

The topological defects of the graphene layers can be interpreted as fractons [18] (both fractons and objects of the Hausdorff dimensionality are identified with the SANS method [21], however, the fracton density of states is a few more complicated).

The Cosserat theory is not dual to a symmetric tensor theory. In this work, we use the geometrical principle of fracton duality to the Cosserat theory with its the asymmetric stress tensor. In [18] the authors show such a mapping of the momentum and stress tensor to the vector magnetic field and tensor electric field.

In the Cosserat theory [16, 18, 19], the elastic media is supplied with microstructure of microscopic origin, i.e. the local volume element in (2+1)-dimensionality is described by the gapless displacement vector ***u***$_i$ and gapped local orientation θ. In the ground state (*T* = 0), both these fields are constant in the space. It is supposed, the system has global translational and rotation symmetry. To resolve the discrepancy in description of topological defects (singularities in displacement and orientation fields), the dual gauge theory in 1-forms with two gapless and one gapped degrees of freedom has been constructed [18].

Monolayer graphite and graphene themselves have 3-fold rotational symmetry of the hexagonal lattice, at the corners of hexagonal Brillouin zone, there are the neutrality point a Fermi surface. In the graphene structure the in-plane σ bonds are formed from 2$s$, 2$p_x$ and 2$p_y$ orbitals hybridized in a $sp^2$ configuration, while the 2$p_z$ orbital, perpendicular to the layer, builds up covalent bonds [6], σ bonds define rigidity of the graphene sheet, and π bonds give valence and conduction bands. At the honeycomb structure, two dual triangular sublattices are built. At the original lattice, the Fermi points connected to the corners of the hexagonal Brillouin zone. The detail theory from Bloch wave functions, tight-binding spectra of graphite and graphene and cause of using gauge fields are discussed, for example, in [6].

In the dual representation, the total defect density is expressed as rotational defects [18],

$$\rho_{rot} = \rho_{discl} + \rho_\theta = \varepsilon^{ij} \partial_i \partial_j (\phi + \theta) \quad , \tag{1}$$

where $\rho_\theta$ is the defect density specific to the Cosserat theory. The disclination defects have two independent contributions. The asymmetric strain tensor in terms of the angles is

$$\gamma_{ij} = u_{ij} + \varepsilon_{ij}(\phi + \theta) \quad . \tag{2}$$

The local displacements and orientations are separated into regular and singular parts.

From expressions $\rho^i_{disl} = 0$ and $\rho_{disc} + \rho_\theta = 0$ (where $\rho_\theta = \varepsilon^{ij} \partial_i \partial_j \theta$), it follows that a globally defined $u_i$ exists if there are no dislocations and disclinations of $u_i$ are canceled by disclinations of $\theta$.

The action of a dual theory in terms of magnetic and electric fields is given in [18] and, at fixing gauge, may be represented into source-free action with two gapless and one gapped mode.

On the other hand, we can use the corresponding $U(1)$ Hamiltonian in the gapless gauge sector with fractons [23] in the form

$$H = \int d^2 x \frac{1}{2} \left( \widetilde{C}^{ijkl} E^{ij} E^{kl} + B^i B_j \right) \quad , \tag{3}$$

where $E, B$ are the components of electric and magnetic fields. Due to the source-free Gauss law $\partial_i \partial_j E^{ij} = 0$,

$$E_{ij} = \varepsilon_{ik} \varepsilon_{jl} u^{kl} \quad , \tag{4}$$

Using canonical conjugations, in particular $u_i$, and $\pi_i$ variables, the Hamiltonian (3) is replaced to the form

$$H = \int d^2 x \frac{1}{2} \left( C^{ijkl} u^{ij} u^{kl} + \pi^i \pi_j \right) \quad , \tag{5}$$

The partition function of (5) is $Z = \int Du \exp(-H/k_B T)$.

The Monte Carlo simulations for disclination energy on a dual triangular lattice are shown at Fig. 3.

An effective semiempirical elastic potential for graphene performed in DFT calculations ([16 and references therein]) may be used for quantitative measures:

$$E_0 = \frac{3}{16} \frac{\alpha}{d^2} \sum_{ij} \left( r_{ij}^2 - d_{cc^2} \right)^2 + \frac{3}{8} \beta d_{cc^2} \sum_{ijk} \left( \theta_{ijk} - \frac{2\pi}{3} \right)^2 + \gamma \sum_{ijkl} r_{i,jkl}^2 \quad , \tag{6}$$

Here $r_{ij}$ are the distances between two bond atoms, $α$ = 26.060 eV/Å, $β$ = 5.511 eV/Å$^2$, and $γ$ = 0.517 eV/Å$^2$.

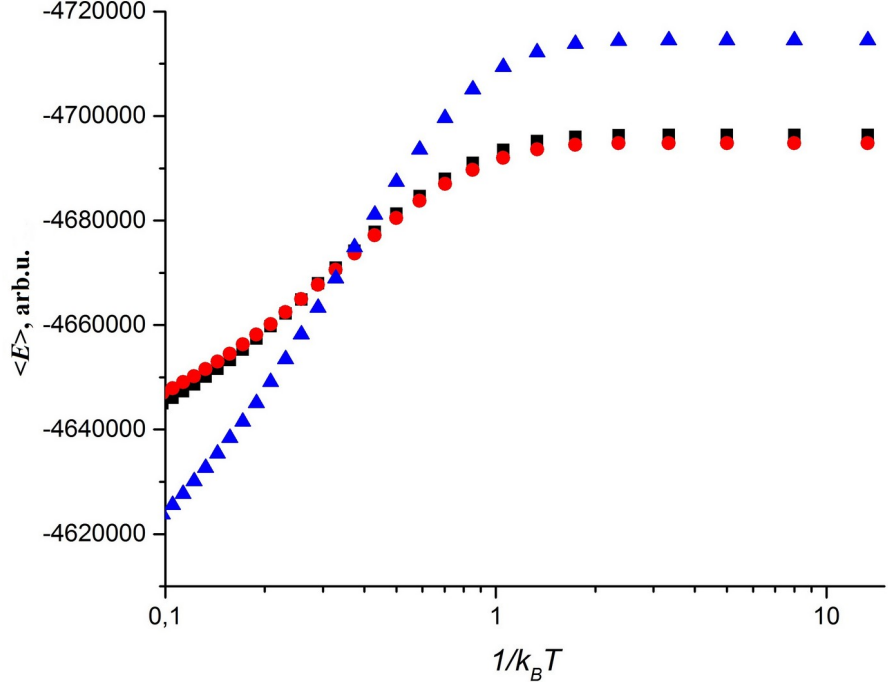

Fig. 3. Temperature dependencies of the average energy of disclinations at different particle concentrations obtained in Monte Carlo simulations with accuracy 0.1%, $N$=86400.

In the Bruggeman-Landauer approximation of the effective media theory, the dielectic properties of a whole composite material can be expressed in the form

$$\frac{\sigma_e - \sigma_1}{2\sigma_e + \sigma_1} p + \frac{\sigma_e - \sigma_2}{2\sigma_e + \sigma_2}(1-p) = 0, \qquad (7)$$

where $p$ is the concentration of the first phase, $\sigma_e$ is the effective electric conductivity, $\sigma_1$ and $\sigma_2$ are the conductivities of two different constituents of a composite. Basing on the measured dielectric values and contributions of phases from (7), we can estimate six mechanical characteristics, the Cosserat elastic constants [16]: Young's modulus, shear modulus, Poisson's ratio, torsion and bending characteristic lengths, Coupling number and polar ratio.

Similar scale dependent electric constant $\varepsilon(r)$ [33] was proposed by Kosterlitz and Thouless on the self-consistent equation, applicable in the BKT theory

$$\varepsilon(r) = 1 + 4\pi^2 y_0^2 K_0 \int_a^r \left(\frac{r'}{a_0}\right)^{4-2\pi U(r')} \frac{dr'}{r'}. \qquad (8)$$

$$K_0 = \frac{1}{2\pi} \frac{\mu_0 B_0}{\mu_0 + B_0} \frac{a_0^2}{k_B T},$$

Here, for the triangular lattice $\mu_0$ and $B_0$ are the shear and bulk moduli in absent of dislocations.

The values $y_0$ (*viz* ln $y_0$) and $a_0$ relate to the core energy and triangular lattice spacing respectively. $U(r')$ is the force of dislocations

4. **Conclusions**

Basing on different methods of characterization morphology of carbon allotrope nanofillers graphite and GNP in the volume of iPP and reconstruction of their shape, we apply evaluating numerical modeling in frames of the Cosserat elastic theory and the fracton representation, which correlates with the SANS results in the available range of nanofiller concentrations. We showed, that the Cosserat elastic theory allows us to provide mapping mechanisms of evolution of disclinations and other topological defects in carbon nanostructures to mechanical characteristics of the polymer nanocomposites.

**Declaration of competing interest**
The authors declare that they have not competing financial interests or personal relationships that could have appeared to influence the work reported in this paper.

**Acknowledgements**
This work was supported by the Ministry of Science and Higher Education of the Russian Federation.

**References**
[1] Polymer Composites. V. 2. Eds. T. Sabu, J. Kuruvilla, M. Sant Kumar, G. Koichi, and S. Meyyarappallil Sadasivan. Wiley-VCH VerlagGmbH Co. KGaA. 2013.
[2] T. Hasan, V. Scardaci, P.H. Tan, F. Bonaccorso, A.G. Rozhin, Z. Sun and A.C. Ferrari (2011) Nanotube and Graphene Polymer Composites for Photonics and Optoelectronics. In: Hayden O, Nielsch K (eds.), Molecular- and Nano-Tubes, Springer, pp 279-354. DOI: 10.1007/978-1-4419–9443-1_9.
[3] V.G. Shevchenko, S.V. Polschikov, P.M. Nedorezova, A.N. Klyamkina, A.N. Shchegolikhin, A.M. Aladyshev, V.E. Muradyan, In situ polymerized poly(propylene)/graphene nanoplatelets nanocomposites: dielectric and microwave properties, Polymer 53 (2012) 5330–5335.
[4] S.V. Polschikov, P.M. Nedorezova, A.N. Klyamkina, A.A. Kovalchuk, A.M. Aladyshev, A.N. Shchegolikhin, V.G. Shevchenko, V.E. Muradyan, Composite materials of graphene nanoplatelets and polypropylene, prepared by in situ polymerization, J Applied Polymer Sci. 127 (2013) 904–911.
[5] P.M. Nedorezova, V.G. Shevchenko, A.N. Shchegolikhin, V.I. Tsvetkova, and Yu.M. Korolev, Polymerizationally filled conducting polypropylene graphite composites prepared with highly efficient metallocene catalysts, Polym. Sci. Ser A 46 (2004) 242–249.
[6] M.A.H. Vozmediano, M.I. Katsnelson, F. Guinea, Gauge fields in graphene, Phys. Rep. 496 (2010) 109–148.
[7] M.A. Milani, D. González, R. Quijada, N.R.S. Basso, M.L. Cerrada, D.S. Azambuja, G.B. Gallan, Polypropylene/graphene nanosheet nanocomposites by in situ polymerization: Synthesis, characterization and fundamental properties, Composites Science and Technology 84 (2013) 1–7.
[8] S.V. Polschikov, P.M. Nedorezova, O.M. Komkova, A.N. Klyamkina, A.N. Shchegolikhin, V.G. Krasheninnikov, A.M. Aladyshev, V.G. Shevchenko, V.E. Muradyan, Synthesis by Polymerization in Situ and Properties of Composite Materials Based on Syndiotactic Polypropylene and Carbon Nanofillers, Nanotech. in Russia 9(3-4) (2014) 175–183.
[9] S.V. Polschikov, P.M. Nedorezova, A.N. Klyamkina, V.G. Krasheninnikov, A.M. Aladyshev, A.N. Shchegolikhin, V.G. Shevchenko, E.A. Sinevich, V.E. Muradyan, Composite Materials Based on Graphene Nanoplatelets and Polypropylene Derived via In Situ Polymerization, Nanotech. in Russia 8(1-2) (2013) 69–80.
[10] S.V. Polschikov, P.M. Nedorezova, A.N. Klyamkina, A.A. Kovalchuk, A.M. Aladyshev, A.N. Shchegolikhin, V.G. Shevchenko, V.E. Muradyan, Composite Materials of Graphene Nanoplatelets and Polypropylene, Prepared by In Situ Polymerization, J. Appl. Polym. Sci. (2012) 1–8. DOI: 10.1002/APP.37837.
[11] L.V. Elnikova, A.N. Ozerin, V.G. Shevchenko, P.M. Nedorezova, A.T. Ponomarenko, V.V. Skoi, A.I. Kuklin, Spatial structure and aggregation of carbon allotrope nanofillers in isotacticpolypropylene composites studied by small-angle neutron scattering, Fullerenes, Nanotubes and Carbon Nanostructures. 29(10) (2021) 783–792.
[12] L.A. Feigin, D.I. Svergun. Structure Analysis by Small-Angle X-ray and Neutron Scattering. Plenum Press, New York, 1987.
[13] A. Gromov, P. Surówka, On duality between Cosserat elasticity and fractions, SciPost Physics 8 (2020) 065-1–065-16.

[14] E. Cosserat, F. Cosserat Theorie des Corps Deformables. Paris.: Hermann *et* Fils. 1909.
[15] J. Li, C. S. Ha, R. S. Lakes, Observation of squeeze twist coupling in a chiral three-dimensional isotropic lattice, Phys. Status Solidi B. 257(10) (2019) 1900140-1-1900140-6. [16] Z. Rueger and R.S. Lakes, Experimental Cosserat elasticity in open cell polymer foam, Philosophical Magazine 96 (2) (2016) 93–111.
[17] O.V. Yazyev, S.G. Louie, Topological defects in graphene: Dislocations and grain boundaries, Phys. Rev. B. 81 (2010) 195420-1-195420-7.
[18] A. Gromov, P. Surówka, On duality between Cosserat elasticity and fractons, SciPost Physics 8 (2020) 065-1–065-16.
[19] P. Surówka, Dual gauge theory formulation of planar quasicrystal elasticity and fractons, Phys. Rev. B. 103 (2021) 201119-1–201119-6.
[20] I.V. Fialkovsky, M.A. Zubkov, Elastic Deformations and Wigner–Weyl Formalism in Graphene, Symmetry 12 (2020) 317-1–317-29.
[21] S. Alexander, C. Laermans, R. Gorbach, and H.M. Rosenberg, Fracton interpretation of vibrational properties of cross-linked polymers, glasses, and irradiated quartz, Phys. Rev. B. 28(1983) 4615–4619.
[22] M. Pretko and L. Radzihovsky, Fracton-elasticity duality, Physical Review Letters 120 (19) (2018) 195301.
[23] M. Pretko, Z. Zhai and L. Radzihovsky, Crystal-to-fracton tensor gauge theory dualities, Phys. Rev. B. 100 (2019)134113.
[24] A.A. Arbuzov, V.E. Muradyan, B.P. Tarasov, Synthesis of Graphene like Materials by graphite Oxide Reduction, Russ. Chem. Bull. 62 (2013) 1962–1966.
[25] W. Spaleck, F. Kuber, A. Winter, J. Rohrmann, B. Bochmann, M. Antberg, V. Dolle, E.F. Paulus, The Influence of Aromatic Substituents on the Polymerization Behavior of Bridged Zirconocene Catalysts, Organometallics 13 (1994) 954–963.
[26] A.I. Kuklin, A.K. Islamov, V.I. Gordeliy, Scientific reviews: Two-detector system for small-angle neutron scattering instrument, Neutron News 16(3) (2005) 16–18.
[27] A.I. Kuklin, A.D. Rogov, Yu.E. Gorshkova, P.K. Utrobin, Yu.S. Kovalev, A.V. Rogachev, O.I. Ivankov, S.A. Kutuzov, D.V. Soloviov, and V.I. Gordeliy, Analysis of neutron spectra and fluxes obtained with cold and thermal moderators at IBR-2 reactor: Experimental and computer-modeling studies, Physics of Particles and Nuclei Letters 8.2(2011)119–128.
[28] A.G. Soloviev, T.M. Solovjeva, O.I. Ivankov, D.V. Soloviov, A.V. Rogachev, and A.I. Kuklin. SAS program for two-detector system: seamless curve from both detectors. J of Physics: Conference Series 848(2017) 012020-1–012020-7.
[29] M.V. Petoukhov, D. Franke, A.V. Shkumatov, G. Tria, A.G. Kikhney, M. Gajda, C. Gorba, H.D.T. Mertens, P.V. Konarev, D.I. Svergun, New developments in the ATSAS program package for small-angle scattering data analysis, J. Appl. Cryst. 45(2012) 342–350.
[30] E.A. Kochetov, V.A. Osipov, P. Pincak, Electronic properties of disclinated flexible membrane beyond the in extensional limit: application to graphene, J. Phys.: Condens. Matt. 22 (2010) 395502-1–395502-11.
[31] A. Manta, M. Gresil, C. Soutis, Predictive Model of Graphene Based Polymer Nanocomposites: Electrical Performance, Appl. Compos. Mater. 24(2017) 281–300. DOI: 10.1007/s10443-016-9557-5.
[32] D. Stauffer, A. Aharoni, Introduction to Percolation theory, Taylor & Francis: London. 1994.
[33] M. Peach and J.S. Koehler, The Forces Exerted on Dislocations and the Stress Fields Produced by Them, Phys Rev. 80(3) (1950) 436–439.